\documentclass[11pt]{article}
\usepackage{graphicx}
\title{\bf About the Power Spectrum Of Primordial Gravitational Waves}
\author{Taimur Mohammadi,\footnote{Department of Physics,
University of Kurdistan, P.O.Box 66177-15175,
        Sanandaj, Iran. Email: t.mohammadi@uok.ac.ir}\\ Behrooz Malekolkalami,\footnote{Department of Physics,
University of Kurdistan, P.O.Box 66177-15175,
        Sanandaj, Iran. Email: b.malakolkalami@uok.ac.ir}\\
 and  Khabat Ghamari\footnote{Kurdistan Meteorological Office. Email: xabatq@gmail.com}\\
       }
\begin{document}
\date{\today}
\maketitle
\begin{abstract}
The primordial gravitational waves (\textbf{PGW}) have been generated by inflationary amplification of the primordial (quantum) fluctuations. It is true that they have not been recorded directly so far, but their spectrum can help a lot in solving the basic puzzles of the early universe as Inflation (high) energy scale. In the present work, we give a straightforward method to calculate the spectral energy density (\textbf{SED}) of the relic gravitons   different from that used in e. g. ~\cite{ mirza04,Latham 2005, Yuki 2006}. In our approach, the evolution equations are in terms of the scale factor (instead of conformal time) through  the Lagrange formalism (instead of the transfer function). The presence of the Hubble parameter allows to calculate the power spectrum in the different dynamical regimes.
\end{abstract}

\medskip
{\small \noindent
\hspace{5mm}
\bigskip

\section{Introduction}
\indent
The gravitational waves (\textbf{GW}) are spacetime curvature disturbances which propagate at  speed of light. One of the familiar sources is rapidly accelerated  massive objects which make ripples in their surrounding spacetime.  These waves  were first predicated by Einstein as approximative solutions of the field equations of General
Relativity in 1916~\cite{GWs 2016}. In recent decades, study of  GW  becomes an  interesting topic in theoretical  and experimental physics such  as GW produced during inflationary period or the compact binary system~\cite{Luc Blanche, mirza04}. The first direct detections of GW was from the merging of black holes and neutron stars  reported by the LIGO/VIRGO~\cite{Abbott 2016, Abbott 20171, Abbott 2017}.\\
GW can give  information from regions of spacetime of universe that are not accessible by other source of information as electromagnetic radiation.
From the cosmological standpoint, the GW  play a potential role to extract or infer the information  about theories as Cosmic Inflation. This is due to the tensorial nature of GW,  because some information  cannot be extracted from the scalar perturbations.\\
In this work, we are interested in studying and analysing the GW originating in the early universe. To understand and describe of its physical conditions  of the early universe, the inflationary paradigm is a very successful theory. This theory was welcomed due to solving the major cosmological problems, because it is well--supported by observational data such as Large Scale Structure of the present universe (this theory is well--described  by  the scalar perturbations). On the other hand, the tensorial perturbations predict the existence of a uniform background of gravitational radiation in the universe. Despite the very low intensity of GW, observational evidences and technological  tools have succeeded in detecting them,  the hopes for future are that,  by improved technologies one should be able to test the prediction from inflation which this would be a huge success in confirming the theory.\\
During the inflationary epoch, the other possible physical processes and matter--field fluctuations  are also involved in the generation of GW and can be considered as sources of the wave production. These processes can cause the noise in the wave propagation  which makes a random character for the GW (what is known as stochastic background GW). The study and investigation of  such a background can help us to understand and explore the early universe events  and high--energy physics. In fact, the energy scale for which the inflation can occur is one of the main challenge for theoretical and experimental physics and in this way the GW generated at that time  can play an important role to  probe  the such energy scale. In fact, it is believed that since there are no gravitational waves found at the 5 percent level, thus the energy scale of an inflationary epoch was below the Planck scale.\\
In general case when one deals with  a stochastic  (radiation) signal,  the signal hasn't a meaningful phase data information (due to the random noise effects) and it  is common to use the spectral methods. The spectral energy density (SED)is an important quantity in analysis of  the (periodic) signals to extract information. It identifies at what frequency (or frequencies) is the signal power or, what form does the energy density function over frequency have?. Therefore, to study and analysis of the primordial GW, the power spectrum  can be used  as an useful tool for extracting the information which can not be read from the time domain of the wave. Indeed, these information (as magnitudes of the frequencies components) can be directly related to the energy scale of inflation.\\
In the present article, we  deal with the tensorial perturbations amplified by inflation leading to primordial GW,  by  special focus on their power spectrum. We calculate the spectral energy in a different way from the one used in the previous works, e. g. ~\cite{mirza04, Latham 2005, Yuki 2006}. SED is obtained directly  in terms of scale factor, from the beginning of the universe until  the present age. In the mentioned works, the spectrum  is obtained
for each dynamical regime of the universe (that is radiation, matter--radiation and matter regimes) individually and then patch them together. The corresponding spectrum diagrams are also plotted which through them some physics  are deduced.\\
The work is organized as follows:

In section 2, we obtain the wave equation  governing the perturbations  in terms of the scale factor using the Lagrangian formalism. In section 3,  the power  spectrum of gravitational waves and  corresponding diagrams are presented. Conclusions are given in section 4.
\section{Perturbations in \textbf{FRW} Background}
In the beginning of this section, let's  give a brief overview of the time evolution of gravitational perturbations, needed for the work.
Consider a perturbed FRW spacetime  whose line element can be written in the form~\cite{ mirza04, Latham 2005}
\begin{equation}\label{N7}
ds^2=g_{\mu\nu}dx^{\mu}dx^{\nu}=\left(\bar{g}_{\mu\nu}+h_{\mu\nu}\right)dx^{\mu}dx^{\nu} = a^{2}\left\{-d\tau^2+ \left[\delta_{ij}+h_{ij}(t,x)\right]dx^i dx^j\right\}
\end{equation}
where $\tau$ is the conformal time,  $\bar{g}_{\mu\nu}= diag\{-a^{2}, a^{2},a^{2},a^{2}\}$ is the unperturbed FRW background metric and $h_{\mu\nu}=h_{\mu\nu}(x)$ are the perturbations  satisfying $|h_{\mu\nu}|<<1$,\hspace{2mm}$h_{00}=h_{0i}=0$ and traceless ($h^i_{i} = 0)$ and transverse $(h^j_{i,j} = 0)$ conditions.
The  linearization of  Einstein equations (in presence of an isotropic and perfect fluid)  for the spectrum of perturbations lead to the following equation~\cite{mirza04, Yuki 2006}:
\begin{equation}\label{C00}
h^{''}_{k}+2\left(\frac{a^{'}}{a}\right)h^{'}_{k}+k^{2}h_{k}=0,
\end{equation}
where $h_k=h(\tau,k)=\int d^{3}\textbf{x} e^{-i\textbf{k}\cdot\textbf{x}}h(\tau, \textbf{x})$ is the Fourier transform of perturbations and prime denotes derivative with respect to the conformal time $\tau$. For  later purposes, we obtain  equation (2) by the Lagrangian  formalism in the following discussion.
The gravitational action describing the tensor perturbations is given by\cite{Latham 2005}:
\begin{equation}\label{C00}
S =\int d\tau d\textbf{x}\sqrt{-\bar{g}}\left(R+\frac{1}{2}\Pi_{ij}h_{ij}\right)=\int d\tau d\textbf{x}\sqrt{-\bar{g}}\left(\frac{-\bar{g}^{\mu\nu}}{64\pi G } \partial_{\mu}h_{ij}\partial_{\nu}h_{ij} +\frac{1}{2}\Pi_{ij}h_{ij}\right),
\end{equation}
where $R$ is the Ricci scalar, $\Pi_{ij}$ is the anisotropic stress tensor   and $\bar{g}$ is the determinant of $\bar{g}_{\mu\nu}$\rlap.\footnote{Since, the first order of $h_{ij}$ are considered and also $h_{ii} = 0$, then $ \sqrt{-g} = \sqrt{-\bar{g}}$ .} For  an isotropic and perfect fluid ($h_{ij}(x)=h(x)$ and $\Pi_{ij}= 0 \hspace{2mm} \cite{Steven 2003}$), the action (3) reduces to
\begin{equation}\label{C00}
S=\int d\tau d\textbf{x}\sqrt{-\bar{g}}\left(\frac{\bar{-g^{\mu\nu}}}{64\pi G } \partial_{\mu}h\partial_{\nu}h\right)=\int d\tau d\textbf{x}\mathcal{L},
\end{equation}
which gives the Lagrangian as
\begin{equation}\label{C00}
\mathcal{L} =\frac{\bar{-g^{\mu\nu}}}{64\pi G } \partial_{\mu}h\partial_{\nu}h  \sqrt{-\bar{g}}.
\end{equation}
This Lagrangian is independent of the perturbation $h(x)$, then the Lagrange equations
\begin{equation}\label{C00}
\frac{\partial \mathcal{L}}{\partial h}-\partial_{\mu}\left(\frac{\partial \mathcal{L}}{\partial(\partial_{\mu}h)}\right) =0,
\end{equation}
reduce to
\begin{equation}\label{C00}
\partial_{\mu}\left(\sqrt{-\bar{g}}\frac{\partial \mathcal{L}}{\partial(\partial_{\mu}h)}\right)=0,
\end{equation}
which by substituting (5),  we get
\begin{equation}\label{C00}
\partial_{\mu}\left(a^{4}{\bar g^{\mu\nu}}\partial_{\nu} h(\tau, \textbf{x})\right)=0.
\end{equation}
By inserting the (inverse) background metric in the last equation and taking  the  Fourier transform on both sides (8), one gets
\begin{equation}\label{C00}
h^{''}(\tau, k) + 2\left(\frac{a^{'}(\tau)}{a(\tau)}\right)h^{'}(\tau, k)+ k^{2}h(\tau, k)=0,
\end{equation}
where is the same equation (2). Equation (9) has two independent variables (that is $a(\tau), h(\tau, k)$) and therefore to solve it, one first have to specify  the scale factor (as known time function)  for each evolutionary stage of the universe and matching the solutions at the epoch of the
transition between the different stages. Instead of this method, we replace the derivative with respect to conformal time by derivative with respect to scale factor, we come in to an independent equation. In other words, if  the time dependence is replaced by the scale factor dependence through the familiar replacement $d\tau=\frac{dt}{a}$, then equation (9) takes the following form
\begin{equation}\label{C00}
a^{4}H^{2}h_{aa}(a, k) + \left(4 a^{3}H^{2}+\frac{a^{4}}{2}\frac{dH^{2}}{da}\right)h_{a}(a, k)+ k^{2}h(a, k)=0,
\end{equation}
where $h_{a}(a, k)=\frac{dh(a, k)}{da}$,  $h_{aa}(a, k)=\frac{d^{2}h(a, k)}{da^{2}}$ and  $H=\frac{\dot{a}}{a}$ is the Hubble parameter.  Since the Hubble parameter can be usually described as function of scale factor, that is $H=H(a)$, then equation (10) describes the evolution of perturbations in terms of the scale factor. It is true that the  above equation   has a relatively  more complex appearance than the  equation (9), but as mentioned above, it is an autonomous equation which by  specifying the Hubble rate function $H(a)$, one can proceed to solve it exactly or numerically.\\
\section{The Power Spectrum}
In this section, we are going to compute SED corresponding to the perturbations satisfying the equation (10). This is done by numerical instructions without restoring to the direct solution of (10). In the first step, we should specify  the Hubble parameter as function of scale factor. This parameter as expansion rate in terms of present--day measurable quantities is  given by the Friedmann--Lemaître equation:
\begin{equation}\label{C00}
H^{2}(a)=H_{0}^{2}\left(\Omega_{r}a^{-4}+\Omega_{m}a^{-3}+\Omega_{k}a^{-2}+\Omega_{\Lambda}\right),
\end{equation}
where $\Omega_{r}=9.4\times10^{-5}\simeq 10^{-4}$, $\Omega_{m}=0.3$, $|\Omega_{k}|\leq 0.01$ and  $\Omega_{\Lambda}=0.7$ are radiation, matter, curvature and  dark energy density parameters, respectively. For a flat background $\Omega_{k}=0$, thus (11) reads
\begin{equation}\label{C00}
H^{2}(a)=H_{0}^{2}\left(\Omega_{r}a^{-4}+\Omega_{m}a^{-3}+\Omega_{\Lambda}\right).
\end{equation}
In the second step, we introduce quantities necessary to describe the relationships between perturbations and their corresponding spectrum. The first quantity is the spectral amplitude $\Delta_{h}^{2}(\tau,k)$ defined by
\begin{equation}\label{N7}
<h_{ij}(\tau,\textbf{x})h^{ij}(\tau,\textbf{x})>=\int\frac{dk}{k}\Delta_{h}^{2}(\tau,k),
\end{equation}
which can be written in the reverse form as
\begin{equation}\label{N7}
\Delta_{h}^{2}(\tau,k)=\frac{2k^{3}}{2\pi^{2}} <|h(\tau, k)|^{2}>,
\end{equation}
or, in terms of scale factor
\begin{equation}\label{N7}
\Delta_{h}^{2}(a,k)=\frac{2k^{3}}{2\pi^{2}} <|h(a, k)|^{2}>.
\end{equation}
The spectral amplitude relates the spectral distribution of the amplified fluctuations and
the cosmological kinematic parameters. Indeed, in the two end of inflation time interval, one deals with the amplified modes in  outside and inside horizon and the spectral amplitude  is useful to describe the distribution of the  modes in outside the horizon.\\
The second quantity is SED that introduced by
\begin{equation}\label{N7}
\Omega_{h}(a,k)=\frac{1}{\rho_c}\frac{d\rho}{d\ln k },
\end{equation}
where $\rho$  and $\rho_c$ are energy density and critical energy density respectively. Since, the relic GW (amplified by inflation) with mode inside horizon should be still present today,  they must be accessible to direct observations (by the high improved gravitational detectors). SED characterises the spectrum of the relic waves and thus it is useful to discuss the possible of their direct detection. For the mode inside horizon, SED is related to the spectral amplitude through the relatively simple relation, that is
\begin{equation}\label{N7}
\Omega_{h}(a,k)=k^2\frac{\Delta_{h}^{2}}{12\hspace{0.5mm}a^2H^{2}(a)}=\frac{k^5}{12\pi^{2}}\frac{<|h(a, k)|^{2}>}{\hspace{0.5mm}a^2H^{2}(a)},
\end{equation}
finally, to  obtain the spectral energy   for the present time, by substituting the (conventional) present value of scale factor ($a(\tau_0)=a_0=1$), one gets
\begin{equation}\label{N7}
\Omega(k)=\Omega_{h}(1, k)=\frac{k^5}{12\pi^{2}}\frac{<|h(1, k)|^{2}>}{\hspace{0.5mm}H_0^{2}}=\frac{k^5}{3\pi^{2}}\frac{|h(k)|^{2}}{\hspace{0.5mm}H_0^{2}},
\end{equation}
where $|h(k)|^{2}=\frac{1}{2}\left(\langle |h_{+}|^{2}\rangle+\langle |h_{\times}|^{2}\rangle\right)=\frac{1}{4}\langle h^{ij} h_{ij}\rangle=\frac{1}{4}<|h(1, k)|^{2}>$. The latter relations mean that the contributions of the two polarization states ($+,\times$) of GW is taken to be equal\cite{Spergel 2003}.\\
Let's remember that discussion  and investigation of the equation (18)  is conditional on obtaining the perturbation $h(k)$ satisfying (10) which is generally not exactly solvable. Thus we have to use and  implement the numerical recipes  which  requires the appropriate initial conditions. Knowing that PGW are originated from inflationary period, the appropriate initial conditions can be considered as
\begin{eqnarray}
h\left(a_{e}, k\right)=1  \\ \nonumber
h_{a}\left(a_{e}, k\right)= 0,\\ \nonumber
\end{eqnarray}
where $a_{e}\simeq 10^{-27}$ is the scale factor at the end of inflation \cite{Scott 2003}.\\
In order to make  more sense spectral density (18), its corresponding diagrams are illustrated in Fig.1. The three (color) graphs shown  in the figure are the spectrum diagrams corresponding to three cases of the Hubble rate function (12). The three cases of the Hubble function  and corresponding spectrum graphs are as follows:\vspace{2mm}
\begin{figure}[!h]
\center{\includegraphics[width=12cm]
{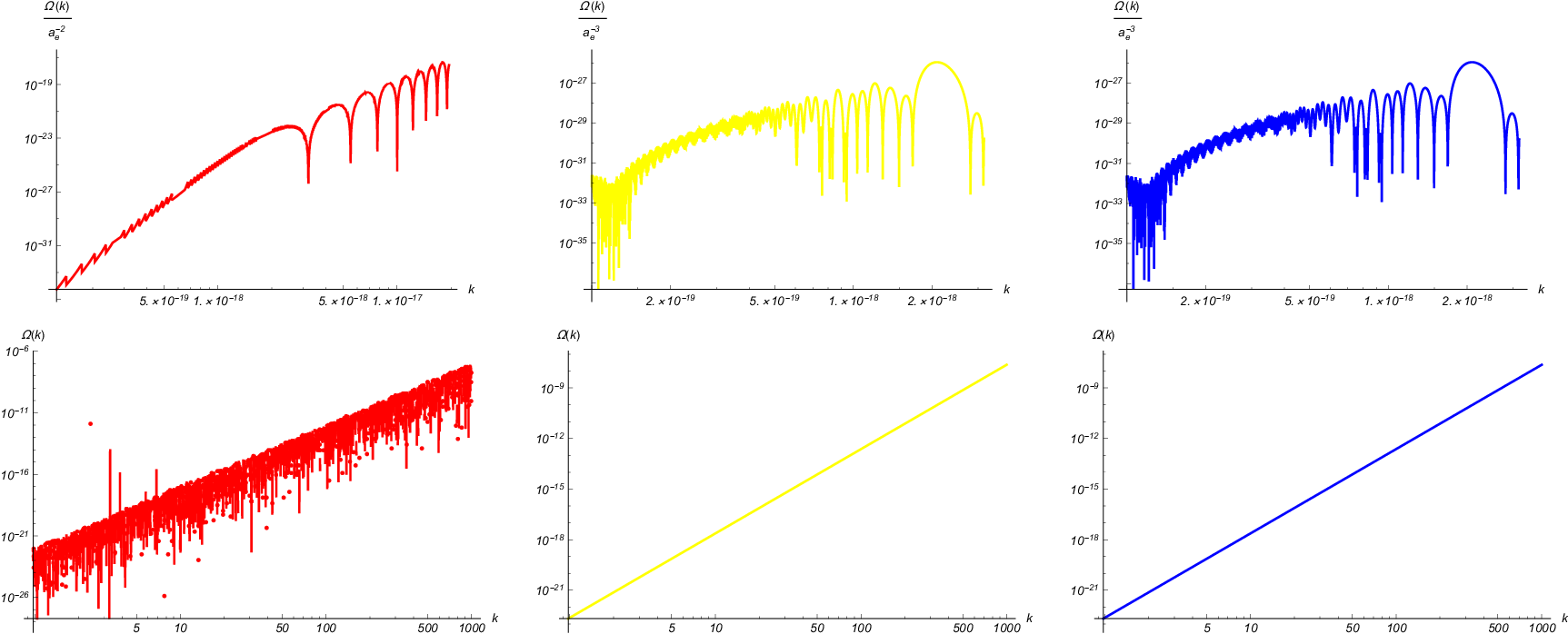}}
\caption{\label{Fig.1}Three  spectrum diagrams of $\Omega(k)$ corresponding to three cases of the Hubble parameter. $H^{2}(a)\approx H_{0}^{2}\Omega_{m}a^{-3}$ (red graph), $H^{2}(a)\approx H_{0}^{2}\left(\Omega_{m}a^{-3}+\Omega_{\Lambda}\right)$ (yellow graph) and $H^{2}(a)=H_{0}^{2}\left(\Omega_{r}a^{-4}+\Omega_{m}a^{-3}+\Omega_{\Lambda}\right)$  (blue graph). The difference between the figures in the top and bottom rows is  only in the  presented frequency ranges.}
\end{figure}
1) General  case: $H^{2}(a)=H_{0}^{2}\left(\Omega_{r}a^{-4}+\Omega_{m}a^{-3}+\Omega_{\Lambda}\right)$ corresponds to the blue graph, \vspace{3mm}\\
2) Matter--Dark Energy dominate case: $H^{2}(a)\approx H_{0}^{2}\left(\Omega_{m}a^{-3}+\Omega_{\Lambda}\right)$ corresponds to the yellow graph, \vspace{3mm}\\
3) Matter dominate case:  $H^{2}(a)\approx H_{0}^{2}\Omega_{m}a^{-3}$ corresponds to the red graph.\vspace{3mm}\\
The two simple can be deduced by looking at the figure:\\
1) All cases are compatible  with the Big Bang Nucleosynthesis (BBN) bounds which states that in frequency band around 100 Hz or $k = \frac{2\pi f}{c} \simeq 628 \hspace{0.5mm} m^{-1}$ (in $c = 1$ unity), $\Omega_{gw}(f)\leq 6.9\times 10^{-6}$ \cite{Abbott 2009}. For a better benchmark, the values of $\Omega$ for this frequncy are given in the Table.1. \\
2) The graphs corresponding to the cases (1), (2) (yellow and blue graphs)  are not much different and this is to be expected due to the insignificant contribution  of radiation term ($\Omega_{r}\simeq 10^{-4}$) in comparison with matter and dark energy contributions.\vspace{1mm}\\
\subsection{Dark Energy Effects}
In this subsection, we discuss impact of  dark energy contribution on spectrum diagrams. In order to this, the diagrams corresponding to matter--dominate (red diagram) and matter--dark energy dominate (yellow diagram) cases are redrawn in Fig.2. The difference with Fig.1 is only in the presented range frequency. The figure says about two effects of dark energy contribution with respect to the matter dominate case:\\
1) It increases SED in the initial frequencies and decreases it in high frequencies.\\
2) It increases the number of power fluctuations in the initial frequencies and at high frequencies, this situation is reversed. (spectrum fluctuations means the number of fluctuations in a frequency interval.)\\
In the end of the section,  we note from an astrophysical  and phenomenological point of view, the allowed constraints on  $\Omega(k)$ (as BBN or observations of millisecond pulsars) can restrict the  values for  Cosmological Constant or other (inflationary) parameters as expansion rate in the inflationary epoch. This depends on the future detections of primordial GW by the new generations of the advanced technological instruments.
\begin{figure}[!h]
\center{\includegraphics[width=12cm]
{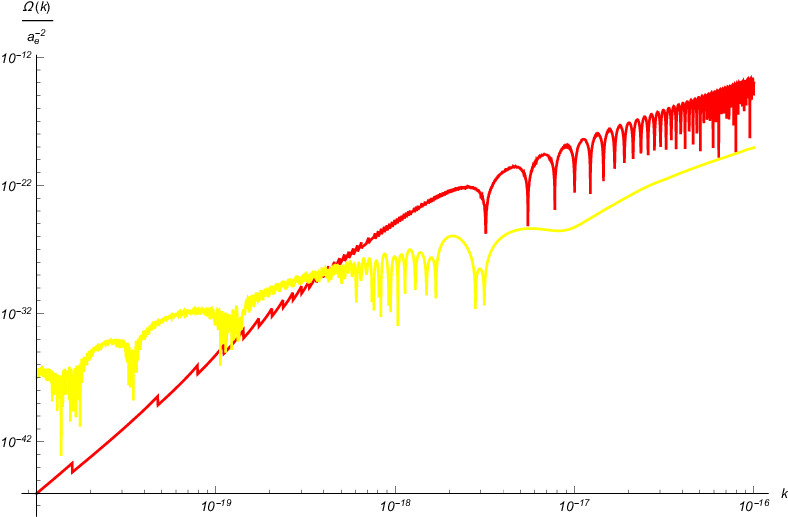}}
\caption{\label{Fig.2}Dark energy contribution effects, $H^{2}(a)\approx H_{0}^{2}\Omega_{m}a^{-3}$ (red graph) and $H^{2}(a)\approx H_{0}^{2}\left(\Omega_{m}a^{-3}+\Omega_{\Lambda}\right)$ (yellow graph) .}
\end{figure}
\begin{table}[h]
\caption{Numerical values of SED at $k\simeq 628$}  
\centering 
\begin{tabular}{c c   } 
\hline\hline 
$\frac{H^{2}(a)}{H_{0}^{2}}$ & $\Omega(k\simeq 628)$  \\ [0.5ex] 
\hline 
$ \Omega_{m}a^{-3}$ & \hspace{2mm}$ 5.83\times10^{-10}$  \\ 
$\Omega_{m}a^{-3}+\Omega_{\Lambda}$ &\hspace{0.1mm} $2.29\times10^{-9} $  \\
$\Omega_{r}a^{-4}+\Omega_{m}a^{-3}+\Omega_{\Lambda}$ & $2.28\times 10^{-9}$ \\ 
\hline 
\end{tabular}
\label{table:nonlin} 
\end{table}
\section{Conclusions}
In the very early universe, tensorial perturbations was amplified during the inflationary epoch leading to primordial radiation. The remnants of this radiation
must exist in the present age as well and thus its detection can tell us the facts about the past history of the universe. An efficient quantity associated with the gravity waves detection is the radiation spectrum (or what is called in breif SED). In this work, SED of primordial gravity waves  is calculated and presented by their corresponding diagrams in three cases of the Hubble expansion rate. The observations (data coming from the phenomenological and
experimental bounds on the spectral energy) predict mostly  growing character for the spectrum and this character is confirmed by the illustrated diagrams here.
The graphs of spectrum are  consistent with the observational data (as BBN bounds) and also indicate the dark energy has a decreasing (in high frequency) and an increasing (in low frequency) effect on the power spectrum.\vspace{2mm}\\
\textbf{Acknowledgments}: We thank K. Rezazadeh for useful discussions.

\end{document}